# UniCalc.LIN:
# a linear constraint solver for the UniCalc system

Evgueni Petrov pes@iis.nsk.su
Yuri Kostov kostov@iis.nsk.su
Elena Botoeva botoeva@ccfit.nsu.ru

Ershov Institute of Informatics Systems SB RAS
Russian Research Institute of Artificial Intelligence
Novosibirsk State University

**Abstract.** In this short paper we present a linear constraint solver for the UniCalc system, an environment for reliable solution of mathematical modeling problems.

## 1 Introduction

The UniCalc system is an environment for reliable solution of mathematical modeling problems. The UniCalc system is based on the constraint programming technology. The constraint solvers which make up the UniCalc system have been used by Russian Research Institute of Artificial Intelligence to efficiently solve design and modeling problems in various areas of industry and science. The first version of the UniCalc system was released in 1990. First of all, the UniCalc system is intended for scientific staff, students, and engineers.

The core solver of the UniCalc system implements subdefinite calculations [1]. For an arbitrary system of constraints (equations, inequalities, Boolean statements), subdefinite calculations find a set containing all solutions (and, possibly, some non-solutions) to these constraints. Besides this core solver, the UniCalc system currently contains a symbolic solver handling the dependency problem which is well-known in the constraint programming community, a symbolic solver for linear equations, and a set of hard-coded strategies describing the typical ways of using these 3 solvers. The user interacts with the UniCalc system through a friendly interface.

Among software for solution of mathematical modeling problems, the UniCalc system is identified by 3 features: reliable calculations, a high-level modeling language, a user-friendly graphical interface. The term "reliable" means here "correct despite rounding errors and inaccuracy in the numerical input data". Another system having these 3 features is the Numerica system [4]; however its support was discontinued in 2003.

Like any constraint propagation algorithm processing real numbers, on some problems, subdefinite calculations converge slowly and/or produce excessively wide enclosures. One can observe this undesirable behavior applying such an algorithm to linear constraints [3]. On one hand, this fact impeded usage of the

UniCalc system for economical modeling because linear constraints are very common in this domain. But on the other hand, this fact motivated us for working on a linear constraint solver for the UniCalc system.

Our work resulted in a reliable (in the same sense as above) solver for linear constraints, called UniCalc.LIN. This solver is integrated into the latest version of the UniCalc system which currently undergoes beta-testing.

In (the full version of) this short paper we present the UniCalc.LIN solver (Section 2), compare it with the symbolic solver for linear equalities and describe strategies of using the UniCalc.LIN solver (Section 3), and discuss our plans concerning solution of linear constraints in the UniCalc system (Section 4).

## 2 The UniCalc.LIN solver

In this section we present the UniCalc.LIN solver for the UniCalc system.

All solvers making up the UniCalc system are required to allow arbitrary mathematical models as the input data and return some mathematical model as the output data. To meet this requirement, the UniCalc.LIN solver performs the following conceptual steps: linear relaxation of the input model, non-reliable calculation of an enclosure for the solutions to thus obtained linear model, adjustment of this enclosure making it reliable. This reliable enclosure is returned to the UniCalc system in the form of linear inequalities each of which contains exactly one variable from the input model.

UniCalc.LIN calculates non-reliable enclosures using a 2-phase lexicographical simplex algorithm which is out of focus of this short paper. We focus on linear relaxation and adjustment of non-reliable enclosures. These two topics are discussed in the following subsections. Note that, for a better performance, these two conceptual steps are merged in the actual implementation.

### 2.1 Linear relaxation

In this section we explain how the UniCalc.LIN solver constructs a linear program which logically follows from the input model. This step is called linear relaxation.

The UniCalc.LIN solver receives the input model in the form of a directed acyclic graph, called syntactic DAG, whose nodes are variables, constants, operations (arithmetic, Boolean, comparison), and the standard mathematical functions.

Each node represents some expression from the input model and cares a reliable enclosure for the range of this expression. For example, if the input model is "$\sin(x + y) = 0$;", then its syntactic DAG consists of the nodes representing $x$, $y$, $x + y$, $\sin(x + y)$, 0, $\sin(x + y) = 0$, and storing some reliable enclosures for the ranges of these expressions.

During linear relaxation, the UniCalc.LIN solver constructs an interval linear program which logically follows from the input model. For each expression $e$ from this model, this program contains the following constraints:

- $e$, if $e$ is a linear constraint;
- $e \in I$, if $e$ is a linear expression with a reliable enclosure $I$;
- $0 = 0$ (no constraint) otherwise.

The constraints of the form "$e \in I$" are rewritten as inequalities; one of the bounds of $I$ may be infinite.

If the linear expressions from the input model contain non-floating-point numbers ($1/10$, $1/3$, etc.), some coefficients of the linear relaxation are intervals.

### 2.2   Reliable enclosure

In this section we explain how the UniCalc.LIN solver finds reliable enclosures for solutions to linear constraints.

Let $C$ be the linear constraints produced by linear relaxation of the input model. If all calculations were exact, it would suffice to solve 2 linear programs for each variable $x$ from $C$: $\min x$ s.t. $C$ and $\max x$ s.t. $C$. The Cartesian product of the intervals of the form $[\min\{x|C\}, \max\{x|C\}]$ would be the wanted reliable enclosure.

Because the exact arithmetic operations on real numbers may consume considerable amounts of memory (and CPU time), the UniCalc.LIN solver uses approximate arithmetic operations on floating-point numbers and the post-processing technique proposed in [2] and adjusted by us for the case of interval linear constraints generated during linear relaxation.

Given a linear program, any reliable enclosure for its solutions and any approximate linear programming solver, this post-processing technique produces another reliable enclosure (hopefully, a smaller one) for the solutions to the same linear program. The diameter of this output enclosure is proportional to the diameter of the input enclosure. The proportionality factor is determined by the accuracy of the approximate linear programming solver: the more accurate solver (smaller residual), the smaller factor. To weaken this limitation, the UniCalc system uses the UniCalc.LIN solver in cooperation with the symbolic solver for linear constraints. This topic is discussed in the next section.

## 3   The UniCalc.LIN solver vs. the symbolic solver for linear equations

In this section we give examples which illustrate weak and strong points of the UniCalc.LIN solver and the symbolic solver for linear equations. We also describe a strategy which increases the chances of using the right linear solver.

The symbolic solver for linear equations is a reliable implementation of Gaussian elimination using interval arithmetic. It solves linear equations very efficiently but may produce excessively wide enclosures if the equations have wide intervals on the right-hand side. Such equations emerge, for example, when inequalities are transformed into equalities by the symbolic solver for the dependency problem.

Consider the following example:

$$0 \le x + y \le 1$$
$$0 \le x - y \le 1$$

The UniCalc.LIN solver alone finds the smallest possible enclosure for the solutions to these inequalities: $x \in [0, 1]$, $y \in [-1/2, 1/2]$.

Cooperation of the symbolic solvers results in a non-trivial enclosure, but not the smallest one. First, the symbolic solver for the dependency problem transforms these inequalities into the following equations:

$$x + y = [0, 1]$$
$$x - y = [0, 1]$$

After that, depending on the order of elimination of variables, the symbolic solver for linear equations produces $x \in [-1/2, 3/2]$, $y \in [-1/2, 1/2]$ ($x$ is eliminated first), or $x \in [0, 1]$, $y \in [-1, 1]$ (otherwise).

Like the symbolic solver for linear equations, the UniCalc.LIN solver may output wider enclosures than possible. This thing happens, if linear equations are ill-conditioned and the enclosure supplied to the UniCalc.LIN solver initially is too loose. Under these conditions the symbolic solver for linear equations still finds relatively small enclosures while the UniCalc.LIN solver cannot make the diameter of the initial enclosure small enough because of a considerable residual in the ill-conditioned equations.

Consider the following example:

$$x + y = 3 \cdot 10^{-7}$$
$$x + (1 + 10^{-7}) = 10^{-7}$$
$$x = y = [-10^7, 10^7]$$

The solution to these equations is $x = 2 + 3 \cdot 10^{-7}$, $y = -2$. The condition number of this linear problem is about $10^7$ (in the sup norm). The UniCalc.LIN solver finds a non-trivial enclosure for this solution: $x \in [1.94, 2.04]$, $y \in [-2.04, -1.96]$ (we round the bounds outward keeping 2 decimal digits). However the symbolic solver for linear equations is much more accurate and finds an enclosure which is almost optimal: $x \in [2.00000029, 2.00000031]$, $y \in [-2.00000001, -1.99999999]$.

The above considerations lead us to the following strategy of using the solvers for linear constraints in the UniCalc system:

- split the linear constraints $C$ obtained by linear relaxation of the input model into linear equations $E$ with thin intervals on right-hand side and other linear constraints $C \backslash E$;
- solve $E$ by the symbolic solver for linear equations and denote the output linear constraints by $\mathrm{Gauss}(E)$;
- solve $C \cup \mathrm{Gauss}(E)$ by the UniCalc.LIN solver.

## 4 Conclusion

In this short paper we presented the UniCalc.LIN solver for reliable solution of linear constraints in the UniCalc system. Reliable solution of linear constraints is an important topic which is not paid much attention in the constraint programming literature. However, during exhaustive search, any reliable solver for non-linear constraints "looks" at constraints very locally. In these circumstances, complicated non-linear constraints "seem" almost linear to the solver and, in fact, it is linear constraints that the solver has to handle most of the time when it solves hard non-linear constraints. This observation is the fundamental motivation for our work.

Currently, our work on reliable solution of linear models in the UniCalc system follows two tracks:

- improvement of the linear relaxation step (doing linearization of non-linear constraints, taking into account the min and max operations, products and ratios of known signs);
- experiments with linear relaxations of limited dimension to limit the CPU time consumed by the UniCalc.LIN solver on large linear models.

We thank Russian Research Institute of Artificial Intelligence and Ershov Institute of Informatics Systems of Russian Academy of Science for supporting this research financially.

## NOTE FOR THE REVIEWRS

The final version of this short paper will also include:

- the results of running the linear solvers of the UniCalc system on an extended set of problems;
- a comparison with the Interval Gauss-Seidel method;
- an extended bibliography list.